\documentclass[11pt]{article}
\usepackage{eurosym}
\usepackage{amsmath}
\usepackage{amsfonts}
\usepackage{amssymb}
\usepackage{ulem}
\usepackage{slashed}
\usepackage{color}
\baselineskip 
\advance
\textheight by
\topskip
\makeatletter

\begin{document}

\title{\textbf{Supersymmetry of a different kind}}

 \author{\noindent \textbf{Pedro D. Alvarez}$^{1}$, \textbf{Mauricio Valenzuela}$^{2}$ and \textbf{Jorge Zanelli}$^{1,3}$\\[5pt]
 ${}^{1}${\small Centro de Estudios Cient\'{\i}ficos (CECs), Casilla 1469, Valdivia, Chile}\\
${}^{2}$ {\small Instituto de Matem\'atica y F\'{\i}sica, Universidad de Talca,  Casilla 747, Talca, Chile}\\
${}^{3}$ {\small Universidad Andr\'es Bello, Rep\'ublica 440, Santiago, Chile}}

\maketitle

\begin{abstract}
A local supersymmetric action for a (2+1)-dimensional system including gravity, the electromagnetic field and a Dirac spin-1/2 field is presented. The action is a Chern-Simons form for a connection of the $OSp(2|2)$ group. All the fields enter as parts of the connection, that transforms in the adjoint representation of the gauge group. The system is off-shell invariant under local (gauge) supersymmetry. Although the supersymmetry is locally realized, there is no spin-3/2 gravitino, and is therefore not supergravity. The fields do not necessarily form supersymmetric doublets of equal mass, and moreover, the fermion may acquire mass through the coupling with the background geometry, while the bosons --the $U(1)$ field and the spin connection-- remain massless.
\end{abstract}

\section{Introduction}  

Supersymmetry is a remarkable symmetry of the basic laws of nature, at least in two accounts: i) it is the only nontrivial extension of the Poincar\'e group that includes internal symmetries and unifies bosons and fermions; and ii) it has never been observed where it is expected, in high energy physics. Here we point out that the lack of evidence of supersymmetry in the standard model may be due to the fact that it is assumed that the bosonic and fermionic fields transform in a vector representation under supersymmetry. However, if those fields were to transform in an adjoint representation as gauge fields do, the resulting theories could have completely different dynamical features. This would lead, in particular, to supermultiplets composed by matter and gauge fields, with different number of bosons and fermions and different masses. The example we present here is possibly the simplest case, in $2+1$ dimensions and for the supergroup $OSp(2|2)$, but the idea can be easily extended to higher dimensions and to other supergroups.

The fact that fermions and bosons can be arranged in the adjoint (rather than vector) representation in three dimensions is suggested by the curious similarity between the Dirac Lagrangian and the Chern-Simons form, $\bar{\psi}\slashed{\partial} \psi \thicksim AdA$. Moreover, although an abelian connection $A_{\mu}$ and a complex Dirac spinor $\psi$ transform very differently under $U(1)$ gauge transformations, they can be accommodated as part of the same connection: take the $3\times3$ matrix, constructed with the  ``natural" representation $\slashed{A}:= \gamma^{\mu} A_{\mu}$, extended with the column $\psi$ and row $\overline{\psi}$,
\begin{equation}\label{int1}
\mathbb{A}=\left[ 
\begin{array}{cc}
A^{\alpha}{}_{\beta} & \psi ^{\alpha } \\ 
\overline{\psi} _{\beta} & 0
\end{array}
\right] =\left[ 
\begin{array}{cc}
\slashed{A} & \psi  \\ 
\overline{\psi}  & 0
\end{array} 
\right] \; .
\end{equation} 
Then, the gauge transformation $A_{\mu}'= A_{\mu}+ \partial_{\mu}\alpha$, $\psi'=e^{i\alpha}\psi$, and $\overline{\psi}'=e^{-i\alpha} \overline{\psi}$ can be obtained as the gauge transformation of $\mathbb{A}$ as a \textit{nonabelian} connection, $\mathbb{A}\rightarrow \mathbb{A}'= g^{-1}\mathbb{A}g + g^{-1} \slashed{d} g$ , where
\begin{equation}\label{U1}
g(x) =\left[ \begin{array}{ccc}
e^{i\alpha(x)} & 0 & 0 \\
0 & e^{i\alpha(x)} & 0 \\
0&0& e^{2i\alpha(x)}
\end{array}
\right]=\exp [\alpha(x) \mathbb{K}],
\end{equation}
with $\mathbb{K}=i$diag$(1,1,2)$, and $\slashed{d}=$diag$(\gamma^\mu \partial_\mu, 0)$. 

This observation suggests considering the $U(1)$ symmetry as a subgroup of a larger nonabelian group, under which $A$, $\psi$ and $\overline{\psi}$ transform as components of the same connection. If that is the case, then one can expect to find a connection with the generators of that symmetry. Naturally, that symmetry should also include transformations that rotate $A$, $\psi$ and $\overline{\psi}$ into each other and hence, it must be some form of local supersymmetry, whose construction is presented in the next section. 

\section{Connection}   
It might be objected that it is not correct to identify the matrix \eqref{int1} as a connection, since it combines components of the connection 1-form $A=A_{\mu}dx^{\mu}$,  with the 0-forms $\psi$ and $\overline{\psi}$, into a matrix of 0-forms, and therefore $\mathbb{A}$ can not be a connection. We now correct this.\footnote{The combination $\gamma^{\mu} A_{\mu}$ is actually a 1-form in the exterior algebra defined by the antisymmetrized product of gamma matrices, and similarly for the differential operator $\slashed{\partial}$. In what follows we will use the more standard exterior algebra of differential forms.}

A more transparent expression is obtained by writing $\mathbb{A}$ as a linear combination of generators in a $3\times3$ matrix representation for a nonabelian algebra with field coefficients. The smallest graded Lie algebra containing a $u(1)$ and a (complex) supersymmetry generator is $osp(2|2)$. So, a natural \textit{ansatz} is the connection $\mathbb{A}=\mathbb{A}_\mu dx^\mu$, with
\begin{equation} 
\mathbb{A}_\mu=A_{\mu} \mathbb{K} + \overline{\mathbb{Q}}_{\beta} (\gamma_{\mu})^{\beta}_{\,\,\alpha} \psi^{\alpha} + \overline{\psi}_{\alpha} (\gamma_{\mu})^{\alpha}_{\, \, \beta}\mathbb{Q}^{\beta} +\omega^a_{\mu} \mathbb{ J}_{a}  ,
\label{A-mu}
\end{equation}
where $\mathbb{K}$, $\mathbb{Q}$, $\overline{\mathbb{Q}}$, and $\mathbb{J}$ are the generators of $U(1)$, supersymmetry and Lorentz transformations in 2+1 dimensions, respectively.  Here $\omega^a_{\mu}=\frac{1}{2}\epsilon^a{}_{bc}\omega^{bc}_{\mu}$ ($\omega^{ab}= -\epsilon^{abc} \omega_c$) is the Lorentz connection. The  generators are explicitly given by
\begin{eqnarray}
&\mathbb{J}_a= \left[\begin{array}{cc}
 							-\frac{1}{2}\gamma_a  	& {\begin{array}{l}
																		0\\
																		0
																	\end{array}} \\
 								0 \; \; 0					&  0
							\end{array} \right] ,  
\mathbb{Q}^{\alpha}=\left\{ \left[
									\begin{array}{ccc}
									 0 & 0 & 0\\
									 0 & 0 & 0\\
									 1 & 0 & 0
									\end{array} \right] ,
								\left[ \begin{array}{ccc}
									 0 & 0 & 0\\
									 0 & 0 & 0\\
									 0 & 1 & 0
						 \end{array} \right] \right\}, & \nonumber \:
\end{eqnarray}
which, together with $\mathbb{K}$ in (\ref{U1}) and $\overline{\mathbb{Q}}_\alpha=(\mathbb{Q}^{\alpha})^T$, span the $osp(2|2)$ graded Lie algebra \cite{Kramer:1978yx}. The nonvanishing (anti-) commutators of this algebra are,
\begin{eqnarray}
&[\mathbb{J}_a, \mathbb{J}_b]= \epsilon_{ab}{}{}^c\mathbb{J}_c \ , \quad \{\mathbb{Q}^{\alpha},\overline{\mathbb{Q}}_{\beta}\}= -\mathbb{J}_a (\gamma^a)^{\alpha}{}_{\beta} -i\frac{1}{2}\delta^{\alpha}{}_{\beta} \mathbb{K}\, ,& \label{sp}\nonumber\\[5pt]
&[\mathbb{J}_a, \mathbb{Q}^{\alpha}]=\frac{1}{2}(\gamma_a)^{\alpha}{}_{\beta} \mathbb{Q}^{\beta} \ ,\quad 
[\mathbb{J}_a,\overline{\mathbb{Q}}_{\alpha}]= -\frac{1}{2} \overline{\mathbb{Q}}_{\beta} (\gamma_a)^{\beta}{}_{\alpha} \,, &\label{}\nonumber\\[6pt]
&[\mathbb{K},\mathbb{Q}^{\alpha}]= i\mathbb{Q}^{\alpha}\,,\qquad [\mathbb{K},\overline{\mathbb{Q}}_{\alpha}]=-i\overline{\mathbb{Q}}_{\alpha}   
\,.&\label{} \nonumber
\end{eqnarray}  

The connection can be expressed more compactly as
\begin{equation}
\mathbb{A} = A \mathbb{K} + \overline{\mathbb{Q}} \Gamma \psi + \overline{\psi} \Gamma \mathbb{Q} + \omega_a \mathbb{J}^a \label{A}
\end{equation}
where we have explicitly introduced the 1-forms $A=A_{\mu} dx^{\mu}$, $\Gamma:=\gamma_{\mu} dx^{\mu}=\gamma_a e^a_{\mu} dx^{\mu}$, and $\omega^a:= \omega^a_{\mu} dx^{\mu}$. Note that the local frames $e^a_{\mu}$ (dreibein) have been introduced to connect the fermionic fields, defined as spinors on the tangent space, to the base manifold. The geometry of the background, whose 3-metric is  $g_{\mu \nu}(x) = \eta_{ab} e^a_{\mu} e^b_{\nu}$ can be assumed to be fixed, in which case $e^a_{\mu}$ and $\omega^a_{\mu}$ would not be dynamical, as is usually assumed in condensed matter physics. 

\section{Action}      

The Chern-Simons 3-form provides a Lagrangian for the connection $\mathbb{A}$ without additional ingredients. Hence, we define
\begin{equation} \nonumber
L=\langle \mathbb{A}d\mathbb{A} + \frac{2}{3} \mathbb{A}^3 \rangle \, ,
\end{equation}
where $\langle \ \cdots \rangle$  stands for the supertrace, and exterior product of forms is assumed throughout. Using the standard conventions \cite{Notation}, one finds
\begin{equation}
L= 2 A d A + \frac{1}{2}\omega^a d \omega_a +\frac{1}{6} \epsilon_{abc} \omega^a \omega^b \omega^c + \bar{\psi}\Gamma[\overleftarrow{\nabla} - \overrightarrow{\nabla}]\Gamma\psi \,, 
\nonumber
\end{equation}
where $\overrightarrow{\nabla}\equiv d-iA -\frac{1}{2}\gamma_a \omega^a$, and $\overleftarrow{\nabla}\equiv \overleftarrow{d}+iA +\frac{1}{2}\gamma_a \omega^a$.

This Lagrangian is real and explicitly invariant under diffeomorphisms of the base manifold. Its gauge invariance can be easily seen if rewritten as
\begin{eqnarray}
L&=& 2 A d A + \frac{1}{4}[\omega^a{}_b d \omega^b{}_a +\frac{2}{3} \omega^a{}_b \omega^b{}_c \omega^c{}_a ] - 2\overline{\psi} \psi e^aT_a \nonumber \\ 
&& + 2 \overline{\psi} (/\hspace{-8pt}\overleftarrow{\partial}-/\hspace{-8pt} \overrightarrow{\partial} +  2i /\hspace{-7pt}  A + \frac{1}{2} \gamma^a /\hspace{-7pt} \omega_{ab} \gamma^b) \psi |e| d^3x ,
\nonumber
\end{eqnarray}
where $|e|=$det$[e^a{}_{\mu}]=\sqrt{-g}$, $w^a{}_b:=1/2 \epsilon^a{}_{bc} w^c$, and $T^a=de^a+\omega^a{}_{b} e^b$ is the torsion 2-form. 

Apart from the torsional term, this is a standard Lagrangian for a Dirac field minimally coupled to $U(1)$ and $SO(2,1)$ gauge fields. These bosonic gauge fields are, in turn, described by their corresponding Chern-Simons actions. Thus, the invariance of the system under local $U(1)$ and $SO(2,1)$ transformations is straightforward. 

As we show below, under a local supersymmetry transformation, the Lagrangian changes by a total derivative. It is remarkable that this rather ordinary-looking system, complemented by a term that couples the spinor to the background torsion, is locally supersymmetric with gauge superalgebra $osp(2|2)$. Although this supersymmetry is local and the system is invariant under general coordinate transformations, there is no gauging of local translations and hence, no gravitino is required. 

\subsection{Field equations}   

Varying the action with respect to the dynamical fields, yields the following field equations:
\begin{equation}
\delta A: \qquad F_{\mu \nu}=\epsilon_{\mu \nu \lambda} j^{\lambda}
\label{F=psi2},
\end{equation} 
where $j^{\lambda}=-i\overline{\psi} \gamma^{\lambda} \psi |e|$, is the electric current density of a charged spin 1/2 field. Note that this equation does not involve the derivatives of $F$, characteristic of a CS action.\\
Varying with respect to the spin connection yields
\begin{equation}
\delta \omega: \qquad R^{ab} = 2\overline{\psi} \psi e^a e^b. 
\label{R=psi2}
\end{equation}
This equation means that the geometry is conformal to a constant curvature manifold. The matter density $\overline{\psi}\psi |e|$ acts as a dark energy source. Although this is not a constant curvature space, it shares with those geometries the property of being transverse to the vielbein, $R^a{}_b e^b =0$. This implies, in particular that the torsion is covariantly constant,
\begin{equation}
R^a{}_b e^b =DDe^a=DT^a=0.
\label{DT=0}
\end{equation}
Varying with respect to $\bar{\psi}$ yields the Dirac equation,
\begin{equation}
\delta \overline{\psi}: \qquad [\slashed{\partial}  - i \slashed{A} +\frac{1}{2}\kappa - \frac{1}{4}\gamma^a \slashed{\omega}_{ab} \gamma^b +     \frac{1}{2|e|}\partial_\mu(|e|E^\mu_a \gamma^a) ] \psi =0 \, , 
\label{Dirac}
\end{equation}
where $|e|\kappa d^3x \equiv e^aT_a$, and $E^{\mu}_a$ are the inverse dreibein, $E^{\mu}_ae^a_{\nu}= \delta^{\mu}_{\nu}$, etc. Note that $\gamma^a \slashed{\omega}_{ab} \gamma^b=\epsilon ^{abc} \omega_{ab\,\mu} E^{\mu}_c= -e^a_{\,\mu}\omega_{ab\nu}e^b_{\, \lambda} \epsilon^{\mu \nu \lambda} $ defines the coupling between the spinor and the curvature of the spacetime background. 

Although the dreibein was introduced as an auxiliary variable that relates the tangent and the base space, its presence in the Lagrangian is dynamical as much as the other fields. Varying the action with respect to $e^a_\mu$ yields,
\begin{equation}
\overline{\psi}\psi \epsilon^{\mu \alpha \beta} T_{a \alpha \beta} = \left( \overline{\psi}\left[\overleftarrow{\partial_\lambda}  -  \overrightarrow{\partial_\lambda} + 2iA_\lambda \right] \gamma^b \psi \right)  \Delta^{\lambda \mu}_{ba},
\label{T}
\end{equation}
where $\Delta^{\lambda \mu}_{ba}\equiv (E^\lambda_b E^\mu_a-E^\lambda_a E^\mu_b)|e|$, and the derivatives act on $\psi$ and $\bar{\psi}$ only.

\section{Solutions}    

We now survey a few simple, physically reasonable, solutions of the coupled system of equations (\ref{F=psi2}-\ref{T}).

\subsection{Bosonic vacua}      
A purely bosonic configuration ($\psi=0$) implies locally flat connections $A$ and $\omega^{ab}$. Depending on the topology of spacetime, this may allow for nontrivial vortex-like $U(1)$ configurations. Moreover, $R^{ab}=0$ still admits a non flat metric, as can be seen by splitting the spin connection as
\begin{equation}
\omega^{ab} \equiv \overline{\omega}^{ab} + \kappa^{ab},
\label{omega}
\end{equation}
where $\overline{\omega}^{ab}$ is torsion-free and $\kappa^{ab}$ is the contorsion,
\begin{equation}
de^a+\overline{\omega}^a_{\,b} e^b=0, \qquad T^a=\kappa^a_{\,b} e^b.
\end{equation}
Consequently, the curvature splits into the Riemann tensor for the metric, $\overline{R}^{ab}$, plus torsional pieces,
\begin{equation}
 R^{ab}=\overline{R}^{ab}+ \overline{D}\kappa^{ab}+\kappa^a_{\,c}\kappa^{cb}.
 \label{Rab}
\end{equation}
Let us now turn to the torsion. The covariantly constant torsion condition (\ref{DT=0}), are three equations for the nine components of $T^a_{\mu \nu}$. By Lorentz rotations, two components of a Lorentz vector can be eliminated, and by general coordinate transformations can be used to eliminate other three. Thus, the four remaining components could be parametrized as 
\begin{equation}
T^a=\tau \epsilon^{abc} e_b e_c +\beta e^a,
\label{T-ansatz}
\end{equation}
where $\tau(x)$ and $\beta(x)$ are a zero- and a one-form, respectively, both of them Lorentz scalars.\footnote{The function $\kappa$ in the Dirac equation (\ref{Dirac}) can be identified as $6\tau$.}  Substituting this expression in (\ref{DT=0}), leads to 
\begin{equation}
d\tau +\tau \beta=0,  \qquad d\beta=0,
\end{equation}
which imply that either $\tau=0$ and $\beta$ is any closed 1-form, or $\tau\neq0$ and $\beta=-d[log(\tau)]$. As the first case is contained as a limit of the second,\footnote{For instance, if $\tau=\epsilon \exp[f(x)]$ in the limit $\epsilon\rightarrow 0$.} we take
\begin{equation}
T^a=\tau \epsilon^{abc} e_b e_c -\frac{d\tau}{\tau} e^a .
\label{T-ansatz'}
\end{equation}
Note that the normalization of $e^a$ is arbitrary. Indeed, the dreibein enters in the connection $\mathbb{A}$ only through the combination $e^a_{\mu}\psi$, and therefore, the action is insensitive to the local rescaling
\begin{equation}
e^a(x)\rightarrow \tilde{e}^a(x)=\lambda(x) e^a(x), \qquad \psi(x) \rightarrow \tilde{\psi}(x)=\frac{1}{\lambda(x)} \psi(x). \label{rescaling}
\end{equation}
Consequently, the torsion rescales as
\begin{equation}
T^a\rightarrow \tilde{T}^a=\lambda [T^a+\lambda^{-1} d\lambda e^a]\;.
\end{equation}
Substituting (\ref{T-ansatz'}) in this expression, the term linear in $e^a$ can be eliminated by rescaling the vielbein in 
(\ref{T-ansatz'}) by $\lambda=\alpha \tau$, where $\alpha$ is an arbitrary constant. Then, in the rescaled dreibein, the torsion is
\begin{equation}
\tilde{T}^a=\alpha \epsilon^{abc} \tilde{e}_b \tilde{e}_c.
\label{T-ansatz''}
\end{equation}
We conclude that it is always possible to rescale the dreibein in so that the torsion is of the form 
\begin{equation}
T^a=\tau \epsilon^{abc} e_b e_c \; , \qquad \tau=constant.
\end{equation}
In particular, this means that in this frame the fermion mass is $m=3\tau$.

Finally, substituting this expression for the torsion in (\ref{Rab}), the metric part of curvature (Riemann tensor) is found to be
\begin{equation}
\overline{R}^{ab} = -\tau^2 e^a e^b,
\label{AdS}
\end{equation}
where we recognize the cosmological constant as $\Lambda=-\tau^2$.

Around a configuration $\psi=0$, an infinitesimal fermionic excitation does not change the background geometry described by $R^{ab}=0=F$ to first order in $\psi$. The infinitesimal fermionic field satisfies the Dirac equation on a spacetime whose local geometry has constant negative curvature. This is the only propagating local degree of freedom of the theory.

\subsection{Constant curvature solutions}  

Several interesting solutions of (\ref{AdS}) are known. In particular, under the assumption of circular symmetry, the Riemannian part of the geometry can be locally described by metrics in the family\footnote{It should be noted, that these geometries are part of a larger family of solutions, obtained by conformal tranformations (\ref{rescaling}). This symmetry has been found to be of relevance in possible applications to realistic condensed matter systems in \cite{Cvetic-Gibbons}.}
\begin{equation}\nonumber
ds^2 = -f^2dt^2 + f^{-2}dr^2 + (r d\phi - N dt)^2, 
\end{equation} 
where $f^2=(r/\ell)^2-M+(J/2r)^2$, $N=J/2r$, and $\ell=\tau^{-1}$. For $M\ell\geq| J|$ this describes the  black hole solution in 2+1 dimensions (the BTZ spacetime \cite{BTZ}); for $J=0$ and $M=-1$, this is AdS spacetime; for $-|J|<M\ell<0$ the space contains a (naked) conical singularity at $r=0$ corresponding to a spinning point particle \cite{Cone}.

It is well known that for some values of the parameters $(M,J)$, these spaces admit globally defined Killing spinors as described in \cite{Coussaert-Henneaux} for black holes and in \cite{Cone} for conical singularities. Interestingly, in the vacuum case $A=d\phi$, which respects circular symmetry but allows for a magnetic flux at the center, also admits Killing spinors provided the magnetic flux is quantized \cite{EGMZ}.

The existence of Killing spinors implies that the bosonic BPS vacua are stable in supergravity, as they minimize the energy and remain bosonic under a (reduced set of) supersymmetries. An important question is whether these solutions are also perturbatively stable in the present setting, which is addressed below.\\

\section{Symmetries}  

A gauge transformation infinitesimally close to the identity of $OSp(2|2)$, takes $\mathbb{A}$ into $\mathbb{A}'= g^{-1}\mathbb{A} g  + g^{-1} dg$, where $g(x)=\exp \Lambda(x)$, and $\Lambda = \alpha \mathbb{K} + \overline{\mathbb{Q}} \epsilon - \bar{\epsilon} \mathbb{Q} + \lambda^a \mathbb{J}_a$.  This induces a transformation of the fields given by $\delta_{\Lambda} \mathbb{A} = d{\Lambda} +[\mathbb{A},{\Lambda}]=\delta A \, \mathbb{K} + \overline{\mathbb{Q}} \, \delta(\Gamma\psi) + \delta(\overline{\psi}\Gamma) \, \mathbb{Q} + \delta{\omega_a} \, \mathbb{J}^a$, which translates to the component fields as
\begin{eqnarray}
\delta A&=&d\alpha-\frac{i}{2}(\bar{\epsilon}\Gamma\psi+\bar{\psi}\Gamma \epsilon) \label{delA}\\[5pt]
 \delta(\Gamma\psi)&=&  \overrightarrow{\nabla}\epsilon+i\alpha \Gamma\psi +\frac{1}{2}\lambda^a\gamma_a\Gamma\psi  \label{delpsi}\\[5pt]
\delta(\overline{\psi}\Gamma)&=&-\bar{\epsilon}\overleftarrow{\nabla}-i\alpha \bar{\psi}\Gamma -\frac{1}{2}\lambda^a\bar{\psi}\Gamma\gamma_a \label{delbpsi}\\[5pt]
\delta{\omega^a} &=& d\lambda^a+\epsilon^a{}_{bc}\omega^b\lambda^c - (\bar{\epsilon}\gamma^a\Gamma\psi+\bar{\psi}\Gamma\gamma^a\epsilon).
 \label{delomega}
\end{eqnarray} 
Under these transformations, the Lagrangian changes by a boundary term by construction, $\delta L= d\mathcal{C}^{U(1)}_\alpha +d\mathcal{C}^{susy}_{\bar{\epsilon},\epsilon}+d\mathcal{C}^{Lor}_\lambda$, where,
\begin{eqnarray}\nonumber
\begin{array}{l}
\mathcal{C}^{U(1)}_\alpha = 2\alpha dA \ ,\qquad
\mathcal{C}^{susy}_{\bar{\epsilon}\epsilon}=\bar{\epsilon} \overleftarrow{d} \Gamma\psi + \bar{\psi}\Gamma d \epsilon ,\\[5pt]
\mathcal{C}^{Lor}_\lambda= -\frac{1}{2}\epsilon_{abc}\lambda^a R^{bc} +\frac{1}{2} (d\lambda^a+\epsilon^a{}_{bc}\omega^b\lambda^c) \omega_a \, . 
\end{array}
\end{eqnarray}
In sum, under infinitesimal $Osp(2|2)$transformations, the field $\mathbb{A}$ transforms as expected for a connection and the action changes by a surface term, as a quasi-invariant CS form should. It can also be checked that successive gauge transformations of $\mathbb{A}$ form a closed off-shell algebra, $[\delta_{\Lambda}, \delta_{\Delta}] \mathbb{A}= \delta_{[\Lambda, \Delta]} \mathbb{A}$.

\subsection{Field representation of the superalgebra}  

The explicit changes of the component fields under the action infinitesimal local $U(1)$, $SO(1,2)$ and supersymmetry transformations are:

\textbf{$\bullet\,  U(1)$ transformations, $g_{\alpha} = \exp[\alpha(x) \mathbb{K}]$}
\begin{equation}
\delta A_{\mu} = \partial_{\mu} \alpha , \quad \delta \psi= i\alpha(x) \psi , \quad \delta \overline{\psi}= -i\alpha(x) \overline{\psi}, 
\end{equation}
and $\delta  \omega^a{}_{\mu}=0=\delta e^a$.

\textbf{$\bullet$ Lorentz transformations,  $g_{\lambda} = \exp[\lambda^a(x) \mathbb{J}_a]$} \\
Eqs. (\ref{delpsi}), (\ref{delbpsi}) determine the transformation laws for the product $\Gamma \psi=e^a\gamma_a \psi$. Since this is not a fundamental spin $3/2$, but a composite of a spin 1 and a spin 1/2, it belongs to a reducible representation of $1\otimes 1/2=1/2 \oplus 3/2$, $\delta_\lambda(\Gamma\psi) = (\delta_\lambda e^a) \gamma_a \psi + e^a \gamma_a (\delta_\lambda \psi)$, with
\begin{eqnarray}
&& \delta_\lambda e^a = \epsilon^a{}_{bc} e^b\lambda^c, \qquad \delta \omega^a = d \lambda^a +\epsilon^a{}_{bc} \omega^b \lambda^c \\
&& \delta_\lambda \psi = \frac{1}{2}\lambda^a\gamma_a\psi, \quad \delta_\lambda \overline{\psi} = -\frac{1}{2} \overline{\psi}  \gamma_a\lambda^a, \quad \mbox{and} \delta A=0 \, .
\end{eqnarray}

\textbf{$\bullet$ SUSY transformations, $g_{\epsilon} = \exp[\overline{\mathbb{Q}} \epsilon(x) - \overline{\epsilon}(x) \mathbb{Q}]$}\\
The supersymmetry transformation is obtained from (\ref{delA}-\ref{delomega}) setting $\alpha=0=\lambda$. In particular, we will assume $\delta_{susy}(\gamma_\mu \psi)=\gamma_\mu\delta_{susy} \psi$. That is, we assume that under supersymmetry, the spin $1/2$ parts, $\psi$ and $\overline{\psi}$, transform, while $e^a$ remains invariant. Thus, under supersymmetry, the fields transform as
\begin{eqnarray}
\delta A_{\mu} &=& -\frac{i}{2}(\overline{\psi} \gamma_{\mu} \epsilon + \overline{\epsilon} \gamma_{\mu} \psi)\;,  \\ 
\delta \psi &=&\frac{1}{3} (/\hspace{-6pt}\partial -i /\hspace{-7pt}  A -\frac{1}{2} \omega^a{}_{\mu}\gamma^{\mu} \gamma_a)\epsilon \;,  \\ 
\delta \overline{\psi} &= &\frac{1}{3} \overline{\epsilon} (- \overleftarrow{/\hspace{-6pt}\partial} - i /\hspace{-7pt}  A - \frac{1}{2} \omega^a{}_{\mu} \gamma_a \gamma^{\mu} )\;,   \\ 
\delta \omega^a{}_{\mu} &=& -(\overline{\psi} \epsilon + \overline{\epsilon}\psi)e^a_{\mu} - \epsilon^a{}_{bc} e^b_{\mu} (\overline{\psi}\gamma^c\epsilon - \overline{\epsilon} \gamma^c \psi)\;, \\
\delta e^a_\mu &=& 0 \label{delta-e}\;.
\end{eqnarray}
Note that under supersymmetry $A_{\mu}$ and $\omega$ remain real. 

It is worth noting that the invariance of the dreibein under supersymmetry (\ref{delta-e}) is not expected in the standard form of local supersymmetry, i. e., supergravity. This is due to the fact that the generator of translations is not in the algebra $osp(2|2)$, in agreement with the form of the connection (\ref{A-mu}). Choosing $\delta_{susy}(\gamma_\mu)=0$ allows to obtain a linear representation of supersymmetry acting on the fields, whose consistency implies the appearance of additional conditions, as we show in the next subsection.

\subsection{Spin 1/2 projection and absence of gravitini}   
The condition $\delta e^a=0$ implies that the metric $g_{\mu \nu}=\eta_{ab}e^a_{\mu} e^b_{\nu}$ does not transform under supersymmetry and this explains the absence of gravitini, in spite of being a locally supersymmetric theory. Let us examine more closely the supersymmetry transformation in (\ref{delpsi}). Expressed in the coordinate basis,
\begin{equation}
\delta_{susy}(\gamma_\mu \psi)=\nabla_\mu \epsilon \, .
\label{S-psi}
\end{equation}
On the other hand, from (\ref{delta-e}), the left hand side is $\gamma_\mu \delta_{susy}\psi$, and hence
\begin{equation}
\delta_{susy}\psi=\frac{1}{3}\gamma^\mu \nabla_\mu \epsilon.
\end{equation}
Now, multiplying once more by $\gamma_\nu$ and comparing the two expressions, one concludes that $\nabla_\mu \epsilon$ is in the kernel of the projector $P_\nu{}^\mu =\delta_\nu{}^\mu-\frac{1}{3}\gamma_\nu\gamma^\mu$,
\begin{equation}
P_\nu{}^\mu \nabla_\mu \epsilon=0 \,, \quad P_\nu{}^\mu P_\mu{}^\lambda =P_\nu{}^\lambda .
\label{P-del-epsilon}
\end{equation}
This projector extracts out the spin 3/2 of a two-index (vector-spin) field, while its complement, $\delta_\nu{}^\mu-P_\nu{}^\mu=\frac{1}{3}\gamma_\nu\gamma^\mu$ is the projector of the spin 1/2 component. 

The consequence of the projection is that there are no spin 3/2 components on the right hand side of (\ref{delpsi}): no gravitini. The consistency of this condition is ensured by the fact that the projector $P_\nu{}^\mu$ is invariant under the action of the entire gauge group, and in particular under supersymmetry, because we have assumed $\delta_{susy}(e^a_\mu\gamma_a) =\delta_{susy}\gamma_\mu=0$.

\section{Discussion and summary}  

The adjoint representation arises naturally in gauge theories which, in the case of supersymmetry corresponds to supergravity. Local supersymmetry seems to be the right structure to describe an invariance in a local field theory, where the symmetry transformations are performed independently on the neighborhood of every spacetime point. A gauge theory constructed for a superconnection defines a supergravity that generically brings in a new field, the spin 3/2 gravitino. In the model presented here, this is not the case because we have managed to leave the metric out of the supersymmetry transformation. The consistency of this choice is validated by the fact that since no spin 3/2 is required to close the supersymmetry, the gravitini can be projected out without giving up local supersymmetry.

The question of vacuum stability can be rephrased as whether the vacuum remains invariant under supersymmetry (BPS state). In order to address this question, we need to solve the equation $\delta_{SUSY}\psi =0$, or
\begin{equation}
(\slashed{\partial} -i\slashed{A} -\frac{1}{2} \omega^a{}_{\mu}\gamma^{\mu} \gamma_a)\epsilon=0.
\label{D-epsilon}
\end{equation}
The existence of the field spinor $\epsilon(x)$ guarantees that a vacuum configuration with $\psi=0$ will remain bosonic under supersymmetry. In the system presented here, $\nabla_\mu \epsilon$ must satisfy the additional requirement (\ref{P-del-epsilon}), but it is easy to see that this is the case if $\nabla_\mu \epsilon = \alpha \gamma_\mu \epsilon$ for any constant $\alpha$. Therefore, it is sufficient to solve
\begin{equation}
[\partial_\mu -iA_\mu - \frac{1}{2} \overline{\omega}^a{}_{\mu} \gamma_a- \alpha \gamma_\mu] \epsilon=0,
\label{K-spinor}
\end{equation}
where the contorsion $\kappa^a\gamma_a$ has been combined with the $e^a\gamma_a$ coming from the right hand side.

Direct substitution of $\overline{\omega}^a_{\mu}$ and $e^a_{\mu}$ into this equation, suffices to solve it for some simple backgrounds which are locally ``pure gauge" but globally nontrivial, like $A=nd\phi$; $M=-1$, $J=0$ (anti-de Sitter); $|M|\ell=|J|$ (extremal black holes and naked singularities). It is a simple exercise to check that in all these cases equation (\ref{K-spinor}) admits solutions, as in the case of vanishing torsion and negative cosmological constant \cite{Coussaert-Henneaux}.

The model presented here could be useful as a simple field theory in 2+1 dimensions for a charged spin 1/2 field in the presence of a $U(1)$ potential, interacting with the background geometry. In this sense, it could find applications in some condensed matter systems with fermionic excitations in 2+1 dimensions like graphene, high $T_c$ superconductors, and in the fractional quantum Hall effect. In these contexts, the spacetime geometry can be assumed as a non-dynamical --although not necessarily trivial-- background, provided by the material substrate on which the fermionic excitations propagate. A particularly interesting possibility in this direction has been recently proposed in \cite{Cvetic-Gibbons}. The effects of the interaction between the fermions and the geometry as well as the effective mass contribution, could be compared with those predicted by supersymmetry in the form given here. 

This model may also be regarded as an example that could be extended to higher-dimensional scenarios  and more realistic symmetry groups, for applications to high energy physics. The crucial point is that supersymmetry could manifest itself very differently if the fermion fields are also accommodated in the adjoint representation. 

Chern-Simons actions with local supersymmetries similar to this have been
studied as supergravities in higher (odd) dimensions \cite{Troncoso-Z}. In those
models, the fermionic sector includes one or several gravitini, and in each
dimension the supergroup that extends the AdS or Poincar\'e group is used. That,
gives rise to a very particular family of theories that exhibit gauge
supersymmetry . An essential ingredient for the algebra considered in those
theories, was the requirement that the metric transforms under supersymmetry,
and therefore models like the one discussed here were not considered. 

The fact that the fermion mass can arise as an effect of the background spacetime through the torsion --which in the case of a condensed matter corresponds to the presence of dislocations in the crystal lattice--, seems to have been put forward by Weyl more than 80 years ago \cite{Weyl}. This might also be an interesting light on which to examine the possibility of giving mass to neutrinos without necessarily breaking any symmetry.  Conversely, it is also natural in this model to interpret the fermionic density $\bar{\psi}\psi$ as a contribution to the cosmological constant (e.g., dark energy).

Supersymmetry realized as a local symmetry is appealing for various reasons. First, being an internal symmetry that rotates different field components among themselves, is more naturally understood as being a local symmetry. Rigid rotations among fields throughout the entire universe and for all time seem to violate locality, relativistic invariance and causality, and might be considered as the limiting case of a very slowly varying gauge parameter at best.  Second, the fact that it is a gauge symmetry makes it harder to break. And third, gauge symmetries do not rely on the equations of motion being satisfied or not, and therefore they are more likely to survive in the quantum version of the theory.

The extension to of the system presented here higher dimensions and other gauge groups will be presented in a forthcoming publication \cite{AVZ2}.

\textbf{Acknowledgements}\\
Many interesting discussions with E. Ay\'on-Beato, A. P. Balachandran, F. Canfora, A. Das, J. D. Edelstein, R. Fernandes, J. Gegenberg, G. Giribet, M. Hassa\"{\i}ne, M. Henneaux, L. Huerta, V. Husain, B. Julia, S. Sorella, R. Troncoso, A. Zerwekh, are warmly acknowledged. This work has been partially supported by FONDECYT grants 1100755, 1085322, 1100328, 1110102, 3100140; Anillos de Investigaci\'on en Ciencia y Tecnolog\'{\i}a, projects ACT-56, Lattice and Symmetry, and by the Southern Theoretical Physics Laboratory ACT- 91 grants from CONICYT. The Centro de Estudios Cient\'{\i}ficos (CECs) is funded by the Chilean Government through the Centers of Excellence Base Financing Program of CONICYT.


\begin{thebibliography}{99}
\bibitem{Notation} Our signature conventions are $\eta_{ab} = diag(-,+,+)$, $\epsilon_{012}=1 =-\epsilon^{012}$. The $\gamma$-matrices are in the representation $\gamma^0= i\sigma^3$, $\gamma^1= -\sigma^1$, $\gamma^2=  \sigma^2$.
\end{thebibliography}
\end{document}